\documentclass[12pt,a4paper]{article}
\usepackage{amsmath}
\begin{document}

\title{\Large On the Non-Lorentz-Invariance of {\sc M.W. Evans}' O(3)-Symmetry Law}

\author{Gerhard W.~Bruhn\\
Darmstadt University of Technology\\ 
D 64289 Darmstadt, Germany\\
{\small bruhn@mathematik.tu-darmstadt.de}}


\maketitle

\begin{abstract}
In 1992 {Evans} proposed the $O(3)$ symmetry of electromagnetic 
fields by adding a constant longitudinal  ghost magnetic field to the well-known 
transverse electric and magnetic fields of  circularly polarized plane waves, and
has since then elevated this so-called symmetry to
the status of a {\em new law of electromagnetics}. 
As a law of physics must be invariant under admissible coordinate transforms, 
namely Lorentz transforms,
 in 2000 he published a proof of the Lorentz invariance of $O(3)$ symmetry of 
electromagnetic fields. This proof is incorrect, because it
is 
erroneously premised on the wave number and frequency being invariant under Lorentz transforms. 
The simple removal of this erroneous premise is sufficient
to show that the  $O(3)$ symmetry is not Lorentz invariant, and thus is not a  
valid law of physics. Furthermore, as the $O(3)$ symmetry later
became the basic assumption of Evans' covariant
grand unified field theory (CGUFT), recently renamed as the
Einstein-Cartan-Evans (ECE) theory), this theory is also physically invalid.

\end{abstract}

\vspace*{1cm}


\newpage
\section{{\sc M.W. Evans}' $O(3)$ hypothesis}

The assertion of $O(3)$ symmetry is a central concern of  M.W. Evans' 
research since 1992:  He claims
that the transverse magnetic field
of  circularly polarized electromagnetic plane wave  is 
accompanied by a constant longitudinal field ${\bf B^{(3)}}$, 
the so-called ``ghost field".\\
 
Evans considers a circularly polarized plane electromagnetic wave
propagating along the $z$ axis of a Cartesian coordinate system \cite[Chap.1.2]{Evans 1} . 
Using the electromagnetic phase\footnote{Equations from {Evans}' book on
the so-called
covariant
grand unified field theory (CGUFT) \cite{Evans 1} appear 
with equation labels \cite[(nn)]{Evans 1} in the left margin.}\\

\cite[(38)]{Evans 1} \hspace*{4.8cm} $\Phi = \omega t - \kappa z ,$\\

where $t$ denotes time, $\kappa = \omega /c$ is the free-space wave number,
$\omega$ is the angular frequency, and $c$ is the speed of light in free space, {Evans} describes the wave in terms of his 
complex circular basis \cite[(1.41)]{Evans 1}. The total magnetic field is stated
by him as\\
 
\cite[(43/1)]{Evans 1} \hspace*{3cm} ${\bf B}^{(1)} = \frac{1}{\sqrt{2}} B^{(0)}   
({\bf i} - i {\bf j}) e^{i \Phi} ,$\\

\cite[(43/2)]{Evans 1} \hspace*{3cm} ${\bf B}^{(2)} = \frac{1}{\sqrt{2}} B^{(0)}   
({\bf i} + i {\bf j}) e^{- i \Phi} ,$\\

\cite[(43/3)]{Evans 1} \hspace*{3cm} ${\bf B}^{(3)} = B^{(0)} {\bf k} ,$\\ 
 
where $i=\sqrt{-1}$ and ${\bf i}$, ${\bf j}$, and ${\bf k}$ are the Cartesian
unit vectors.
The total magnetic field satisfies his ``cyclic $O(3)$ symmetry relations"\\

\cite[(44/1)]{Evans 1} \hspace*{3cm} 
${\bf B}^{(1)} \times {\bf B}^{(2)} = i B^{(0)} {\bf B}^{(3)*} ,$\\

\cite[(44/2)]{Evans 1} \hspace*{3cm}
${\bf B}^{(2)} \times {\bf B}^{(3)} = i B^{(0)} {\bf B}^{(1)*} ,$\\

\cite[(44/3)]{Evans 1} \hspace*{3cm} 
${\bf B}^{(3)} \times {\bf B}^{(1)} = i B^{(0)} {\bf B}^{(2)*} .$\\
 
Equation \cite[(43/3)]{Evans 1} specifically {\em defines} {Evans}' ghost field ${\bf B}^{(3)}$,
which is coupled by the relations \cite[(44)]{Evans 1} to the transverse    
components  stated as  \cite[(43/1)]{Evans 1} and  \cite[(43/2)]{Evans 1}.\\ 

Evans' {\bf B Cyclic Theorem} is the statement that the magnetic field
components  \cite[(43/1)]{Evans 1} and  \cite[(43/2)]{Evans 1} of each circularly 
polarized plane wave are accompanied by a longitudinal component \cite[(43/3)]{Evans 1}, 
and that all three components together
fulfil the cyclic $O(3)$ symmetry relations \cite[(44)]{Evans 1}.
Evans considers this $O(3)$ hypothesis as a {\bf Law of Physics}.\\
 

\section{Is the $O(3)$ hypothesis Lorentz-invariant?}

A law of physics must be invariant under admissible coordinate transforms, 
namely Lorentz transforms. A circularly polarized plane wave displays this property 
when described by any observer who is moving with uniform
velocity with respect to an inertial frame $K$ at rest. 
Therefore, {\sc Evans'} $O(3)$ symmetry law 
should be valid in all inertial frames of reference.
Hence, to check the physical validity of {Evans}' $O(3)$ hypothesis, 
we shall apply a longitudinal Lorentz transform to the plane wave as described by Evans
(the ghost field included).\\ 

In an article published in 2000 \cite[ p.14]{Evans 2}, Evans 
provided a proof of the Lorentz 
invariance of the $O(3)$ hypothesis \cite[(44)]{Evans 1} by referring to the invariance of 
the vector potential {\bf A} under Lorentz transforms.
That is a good method obtaining the transformed magnetic field, {\em  if done correctly}.\\
 
The vector potentials of the transverse
components ${\bf B}^{(1)}$ and ${\bf B}^{(2)}$ of the plane wave under 
consideration are given by

\begin{equation} \label{(1)}
\left.\begin{array}{l}
{\bf A}^{(1)} = \frac{1}{\kappa} {\bf B}^{(1)} = 
\frac{1}{\kappa\sqrt{2}} B^{(0)} ({\bf i} - i {\bf j}) e^{i \Phi}\\[8pt]
{\bf A}^{(2)} = \frac{1}{\kappa} {\bf B}^{(2)} = 
\frac{1}{\kappa\sqrt{2}} B^{(0)} ({\bf i} + i {\bf j}) e^{- i \Phi} 
\end{array}\right\},
\end{equation}

while the vector potential of the {\em longitudinal} component ${\bf B}^{(3)}$ is 
\begin{equation} \label{(2)}
{\bf A}^{(3)} = \frac{1}{2} {\bf B}^{(3)} \times (x {\bf i} + y{\bf j}) 
= \frac{1}{2} B^{(0)} (x {\bf j} - y {\bf i}) .
\end{equation}

The Lorentz invariance of the vector potential ${\bf A}^{(3)}$ yields the
Lorentz invariance of 
${\bf B} ^{(3)}$ and the factor $B^{(0)}$ in   \cite[(43)]{Evans 1} and \cite[(44)]{Evans 1}
for {\em longitudinal} Lorentz transforms (i.e. between inertial frames $K$ and $ K'$
such that $K'$ moves relative to $K$ with velocity ${\bf v} \parallel {\bf k}$   
and  $\beta = |{\bf v}|/c$).\\
 
Evans \cite{Evans 2} ignored that $\omega$ and $\kappa$ are {\em not} Lorentz-invariant. 
Under {\em longitudinal} Lorentz transforms, we have the well-known Doppler effect:

\begin{equation} \label{(3)}
\omega' = \sqrt{\frac{1-\beta}{1+\beta}}\  \omega ,\qquad \qquad 
\kappa' = \sqrt{\frac{1-\beta}{1+\beta}}\  \kappa .
\end{equation}

Therefore the invariance of the vector potentials {\em does  not transfer} to the 
transverse components
${\bf B}^{(1)}$ and ${\bf B}^{(2)}$.\\

More importantly here, due to the invariance of ${\bf A}$ we obtain 

\begin{equation} \label{(4)}
{\bf B}'^{(1)}\!\times\!{\bf B}'^{(2)} =
\kappa'^2 {\bf A}'^{(1)}\!\times\!{\bf A}'^{(2)} 
=
\frac{1-\beta}{1+\beta}\ \kappa^2 {\bf A}^{(1)}\!\times\!{\bf A}^{(2)}
=
\frac{1-\beta}{1+\beta}\ {\bf B}^{(1)}\!\times\!{\bf B}^{(2)} ,
\end{equation}

from (\ref{(1)}). That is, the expression ${\bf B}^{(1)}\!\times\!{\bf B}^{(2)}$ 
on the left side of \cite[(44/1)]{Evans 1} is not Lorentz-invariant, whereas
${\bf B}^{(3)}$ on the right side of \cite[(44/1)]{Evans 1} is Lorentz-invariant due to (\ref{(2)}). Equation \cite[(44/1)]{Evans 1} therefore,
if valid in the inertial frame $K$, {\em cannot be valid} also in the inertial frame $K'$. 
Hence, Evans' cyclic $O(3)$ symmetry relations \cite[(44)]{Evans 1} 
are {\em not Lorentz-invariant and so cannot be a Law of Physics}.


\section{Epilog}

After the foregoing sections had appeared on http://arxiv.org, Evans reacted polemically on\\ 
{\footnotesize http://www.atomicprecision.com/blog/2006/12/26/jackson-on-the-lorentz-transformation/}. 
Therein, he suggested that Equation (3.111) of another book of his  \cite{Evans 3}  
should prove the Lorentz-invariance of his ``B Cyclics". \\

This equation is as follows:\\

\cite[(3.111)]{Evans 3}\hspace*{3cm} $B^{(0)\prime} = 
[\frac{1-\frac{v}{c}}{1+\frac{v}{c}}]^{\footnotesize 1/2} B^{(0)} . $ \\

It is a consequence of the Lorentz transform applied to the transverse components 
${\bf B}^{(1)}$ and ${\bf B}^{(2)}$. 
However, just before  \cite[(3.111)]{Evans 3}) Evans stated the transformation rule\\

\cite[(3.110-3)]{Evans 3}\hspace*{3cm} $ {\bf B}^{(3)\prime} = {\bf B}^{(3)} , . . . $  \\

This rule implies that
$|{\bf B}^{(3)\prime}| = |{\bf B}^{(3)}|$ 
which immediately leads to a {\em contradiction} in Ref. \cite{Evans 3} itself,
as on p.5 of that book we find the equation\\ 

\cite[(1.2a)]{Evans 3}\hspace*{3cm}$ |{\bf B}^{(1)}| = |{\bf B}^{(2)}| = |{\bf B}^{(3)}| = B^{(0)} $ \hspace*{2cm} \\ 

that yields \\

\hspace*{3cm} $ B^{(0)\prime} = |{\bf B}^{(3)\prime}| = |{\bf B}^{(3)}| = B^{(0)}$ \\ 

in contrast to \cite[(3.111)]{Evans 3}. Therefore, {Evans} stopped his consideration 
in Ref. \cite{Evans 3} just before recognizing the final 
contradiction he would have arrived at.


\end{document}